# Publicity of the intimate text (the blog studying and publication)

One of the important problems of a modern society – communications. At all readiness of this question both humanitarian, and engineering science, process of transfer and information reception remains in the centre of attention of researchers. The dialogue phenomenon in a network becomes the significant factor of such attention.

The possibility of the user choice, freedom of expression of feelings and requirements, communications and information interchange freedom – define principles of the Open Internet not only specificity of communications in networks, but also genre and stylistics possibilities of the texts providing dialogue.

Today Internet users apply various possibilities of the network, one of the most demanded – a blog. The Internet as an open space of dialogue allows the texts created and functioning in a network, to pass and in fiction sphere. We can observe similar processes concerning a genre of a diary and its genre versions: Network diary (Internet diary) – a blog – and an art (literary) diary. The diary as an auto-documentation genre at the present stage of its development gets from especially intimate conditions on Internet space. The blog phenomenon reflects as potential of publicity of the text of the records, peculiar to a genre, and feature of a modern cultural situation. Popularity of Internet diaries defines interest to them and readers of a fiction with what cases of the publication of blogs, significant both in substantial, and in the art relation are connected. So, publications of the Russian bloggers are known: for example, the artist Peter Lovygin (lovigin.livejournal.com) [3], etc.

In diary records of writers it is possible to see interaction autodocumentary and art has begun: transition diary narrative in a work of art. Similar possibilities of the diary text are defined first of all by the factor of the Author – the writer is focused on publicity of the life and creativity. The tradition of the publication of all texts of the author created by culture including inartistic, leads to comprehension of that the archive, and first of all a diary, will be published. In a modern situation the creative heritage structure joins also network texts.

As a material for research the blog of writer Evgeny Grishkovets is chosen, which author conducted throughout five years within the limits of the Russian version «Livejournal» – «Live Journal», and has published in the form of the story «The Year of LLife» (2008) [1].

Space of crossing art and documentary influences to the communications organization. Dialogue in a blog sphere takes the important place: the considerable part of a blog is given to comments of readers-bloggers and answers to these records. As a result of this dialogue by co-authors there are also readers of a blog. A blog combines properties of various forms of communications in a network (a forum, a chat) as simultaneously is the personal hypertext document created on the basis of structure set by a site, and the interactive tool of speech interaction in the form of a polybroad gull.

Intensions of preservations of freedom of the dialogue distinguishing a blog, are present and at E. Grishkovets's literary diary. Thus, the fact of communication with readers and an exchange of opinions is understood by the author as the basic function of a diary. And the author notes that the relation to content and style of commentaries determines strictly the means of the author of this blog: *I not has so long ago said HERE several words about the foolish and self-reliant commentaries. And separately over-anxious accepted my words to their calculation. This pleases itself me [1].*

Therefore the author both provokes this dialogue, and tries to supervise it according to the representations about the maintenance and style of the text. Systematically the comments showing aggression, the disrespectful relation, excessive criticality left. Such behaviour of Grishkovets explains the desire to create communicative space, comfortable for it and other readers, and also not contradicting its aesthetic and moral principles: *I move away separate commentaries most frequently for that reason, that they are hostile or superficially angry. These are the commentaries of those, who wander along the dark corridors of the Internet with the unkind and indistinct desires. […] Me seems this by the completely polite and tactful method to man the door. But the atmosphere HERE exists, and I to her shore* [1].

Thus in the art text of a documentary component of a diary the author does not refuse representation: the subject line of relations of bloggers corresponds and with a known subject situation the *creative person – crowd*, and archetype interrelation *I am–Another*, and with a concrete situation of dialogue of the known writer and its readers.

Dynamism of an Internet diary is connected not only with the maintenance of records (this reflexion of a tide of life), but also with the dialogue organisation: the text essentially is not fixed. The blog text can constantly replenish both records of the author, and comments of readers. Thus diary existence in a global information field causes absence of direct contact of bloggers and their anonymity. But also defines ethics of behavior in a blog: necessity of respect of the rights and values of participants of communications on the Internet, despite informality of this public communicative situation.

The blog, as well as a diary genre, is extremely subjective: the author addresses to the description of those situations and persons which are interesting to the blogger. Internet diaries of known authors including writers (such is E. Grishkovets's blog) connect as the information, significant only for the author of a blog, and focused on reader's interest, the potential addressee. One of the major possibilities of a blog is a process of formation of personal space and postulation *own I* in interaction with consciousness of *Another*, i.e. in a communicative space of a blog spheres as communities of persons. Grishkovets characterizes his blog as the space of active contact, precisely, the exchange of opinions it forms the special text: *In one-and-a-half years of presence HERE I observed quite strange things. Several generations of the actively inking and commenting on people were replaced. It regularly read messages of the type: "I thought that you good, believed to you, it loved that the fact that you make, and now she understood that you poor, evil, deceitful. I depart from you… "(well, and after such words it most frequently went insincere the wish of successes and prosperity.) [2]*. The motion of text in the space of Internet-page forms special three-dimensional relations in the text, is added topos, described by the author of blog in a word here.

Texts of records in a blog are characterised by a combination of a literary language to loans speaking another language, youth slang, a computer slang. Necessity of transfer of intonational parameters of oral speech defines active use of abbreviations, paralinguistic and graphic means. Similar synthesis characterises also style of a literary diary of E. Grishkovets. So, the author meaningly refuses a popular symbol *a Smilie*, in different variants functioning on language space of a network. In network and further in a literary diary the author consistently uses the note *a smile*. See: *These warnings please themselves me, because me it is desirable to believe that in them is contained sincerely the uneasiness for me (smile). [2]* Only in one case a word *the smile* is replaced with a traditional designation of a symbol in the form of graphic means: *Friends [...] have sent one and only SMS: "As to us now to go here to our restaurants: ("* [1].

On it is underlined literary character of the note specifies both its conclusion in brackets, and communication with note function in a plot. End of a fragment of the text by the note *the smile* shows an emotional condition of the author or its relation to the described events, or the story about a comic situation and aspiration to cause certain reaction of the reader. *These warnings please themselves me,*

*because me it is desirable to believe that in them is contained sincerely the uneasiness for me (smile) [2]*. Compare: *It would write still, but indeed I dictate, and the wife has other matters and furthermore it does not be worthwhile to lay out too long texts. Yes even life is in front long (smile) [1]*. Thereby also the communicative nature of a blog and in the form of the published literary diary is supported.

Connection of an Internet diary with art genres defines also lines of style of the Grishkovets's network text: refusal from сленга, the formulas of politeness focused on traditional forms communications (a letter): *Hello!*, *Your Grishkovets* [2, 4], the critic of hobby for abbreviations (such, as Russian idiom *IMHO* [1]). Thus in the text of a literary diary the part of these signs disappears: instead of the regular greetings opening the message in a blog, in the text of the story there are a Preface and the Epilogue of the author. Structurally these parts of the product are not allocated, but they provide communications of the writer and the reader and are necessary for composite integrity to the story.

Communications in a blog are connected with difficult structure of this phenomenon. Records of the author and comments form is formal and semantic unity where the sequence of references, reactions, comments, answers to the comments, attached to the initial message, creates the ramifying hypertext.

Properties of the hypertext in a blog – nonlinearity, an openness, dynamism, ramification – open frame structure of the network text. Quality of an openness of a blog is shown in expansion of borders of the hypertext. So, the structure of the diary text joins also various references to other Internet sites, and also inclusion of photos, video recordings and audio records.

The hypertext organization of a blog is reflected and in the book structure: Grishkovets gives references to the blog, addresses of pages livejournal.com where the photos and audio materials are stored. Thus and the edition is accompanied by tabs of illustrations that also allows to make certain impression about an Internet diary of the writer.

Popularity of blogs as dialogue forms in a network it is caused by aspiration of the modern person to designate a place of own person, *I am* in a global network. Internet space formation influences perception of network texts: it is not so much way of self-knowledge and a reflexion as it occurs to a traditional diary and the traditional fiction, it is the open and free communications. It defines domination communicative and ascertainment of contacts blog functions. In this connection research of blogs and their interaction with fiction allows to describe process of change of the Internet as social and cultural phenomenon. The fact of the publication of blogs and increasing popularity of bloggers is connected, in our opinion, with an increasing openness of a blog sphere (each record can be commented any user), and accordingly, the Internet as a whole. The generated lines of Internet communications characterise network texts – blogs – as popular, public documents in which private life becomes property of the public.